# CMS Conference Report

# XML for Domain Viewpoints

F. van Lingen[1,2], R. McClatchey[1,4], P. v/d Stok[3], I. Willers[1]

[1]CERN, Geneva, Switzerland

[2]Supported by Eindhoven University of Technology and University of the West of England

[3]University of Technology Eindhoven, Philips Research Laboratories Eindhoven, Prof. Holstlaan 4, 5656 AA Eindhoven, The Netherlands

[4]Centre for Complex Cooperative Systems, Univ. West of England, Frenchay, Bristol BS16 1QY UK

*Abstract*

Within research institutions like CERN (European Organization for Nuclear Research) there are often disparate databases (different in format, type and structure) that users need to access in a domain-specific manner. Users may want to access a simple unit of information without having to understand detail of the underlying schema or they may want to access the same information from several different sources. It is neither desirable nor feasible to require users to have knowledge of these schemas. Instead it would be advantageous if a user could query these sources using his or her own domain models and abstractions of the data. This paper describes the basis of an XML (eXtended Markup Language) framework that provides this functionality and is currently being developed at CERN. The goal of the first prototype was to explore the possibilities of XML for data integration and model management. It shows how XML can be used to integrate data sources. The framework is not only applicable to CERN data sources but other environments too.

Keywords: XML, data integration, mediators, XML queries, domain views

# 1. Introduction

Because of an autonomous approach in the design and construction of the sub-detectors within the CERN Compact Muon Solenoid (CMS) experiment [1] and the different lifecycle phases of the sub-detectors there are numerous types of data sources, and many different interfaces to these sources. In general the data sources are relational, object oriented databases, or flat files that have a pre-defined schema (e.g. XML and XML schema [2]). Retrieving data from multiple sources therefore requires knowledge about the interfaces, the query dialects and the underlying data models. The users (mainly physicists) want to access these data and use them within their own data models.

Currently physicists build hard-coded interfaces to the data they require and transform them to the model (or punch in the numbers by hand) when required. This results in much overhead with respect to building data interfaces since everybody tends to build their own interfaces. Furthermore, punching numbers by hand is error prone and difficult to manage. Because of this physicists are repeatedly confronted with the same computer science problems. Another problem is that it is not possible to manage the interfaces and models that are used for data access since reuse is not promoted. As a result physicists spend less time on the modeling issues and more time on computer science and organizational issues. This process would be simplified if there were a framework that supports the definition of interfaces/models between the different data sources. This would allow physicists to focus more on the modeling issues and promote reuse of other models.

A large number of the data sources relate to the production of the detector (its design, construction, and calibration). With respect to this the problem described here is not unique to CERN. Within large organizations numerous databases can be in use in different departments. Users often need data from these multiple different sources. A solution for this problem could be an ERP (Enterprise Resource Planning) system. ERP systems tend to be monolithic pieces of software. Although they can be distributed, different phases (or groups) in the life cycle have to conform to a global schema in terms of naming and structure. ERP systems do not have the capability to incorporate different data sources without significantly changing them, or copying them into this monolithic system. Furthermore, ERP systems can be difficult to implement [3].

The problem discussed in this paper relates to several areas within research literature including heterogeneous distributed databases [4], [5], [6], mediators [7], [8], and active databases [9]. In [5] and [6] a taxonomy is discussed of different forms of heterogeneity. All these forms are present within the CMS environment. Within this paper we focus on mediator technology based on XML. An important reason for choosing XML is that it is becoming the de facto standard for data exchange/integration, and it is vendor independent. If data would be in XML format or XML compatible it would enable users to pose XML queries over these data [10]. [11] discusses views and XML. A view specification in XML should rely on a data model and a query language as in the relational database world.

The outline of this paper is as follows: The next section discusses the architecture. Section 3 discusses model specifications. Section 4 describes management of models. Section 5 discusses some observations based on the current prototype. Section 6 discusses related work. Finally, section 7 contains the conclusions. Throughout this paper we will use the Xpath notation, and the Quilt query language as examples. Furthermore, we assume that every source represents itself as an XML document. [12] describes how this is achieved for object oriented databases. [27] is a project that focuses on publishing relational data as XML.

Xpath [13] is a navigation mechanism for XML documents. It allows selection of nodes using conditions on the attributes, parent and child nodes. Quilt [14] is a proposal for a query language for XML documents. It is based on Xpath for selecting nodes. But Quilt allows joins over multiple documents and recursive queries and returns the result as an XML document. Its structure is based on features found in query languages like SQL, OQL. In this work we used an implementation of Quilt called Kweelt [15].

# 2. Architecture

Figure 1 shows an architectural overview which we will refer to as a mediator environment. In this architecture the models are separated from the software. The advantage of the separation of software and model, is the reuse of this software for other models. Furthermore, this architecture facilitates management of the models.

Users will issue a query or a set of queries to the mediator environment using an integration model specific for their domain. The integration model is a specification of how data from the sources should be integrated. The model is defined in terms of either other integration models (which can be viewed as virtual sources) or data sources (which are shielded by an XML wrapper).

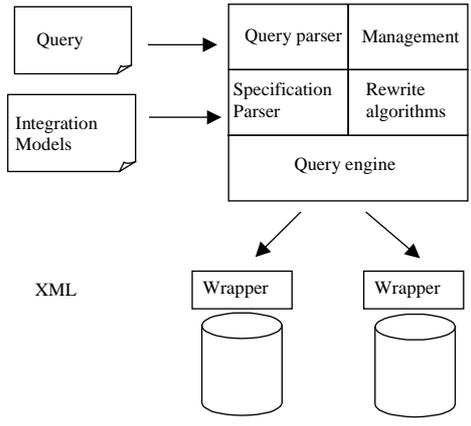

**Figure 1. Architecture**

In the current prototype the model specification contains a description of integration models specified within an XML query language, Quilt. This determines how users' queries will be translated. An integration model is specified as a set of functions (as specified in Quilt). These functions represent data objects and relations between data, for users or other integration models. In order to promote reuse, integration models do not have to be based on sources directly but can be based on other integration models. Figure 2 shows an example.

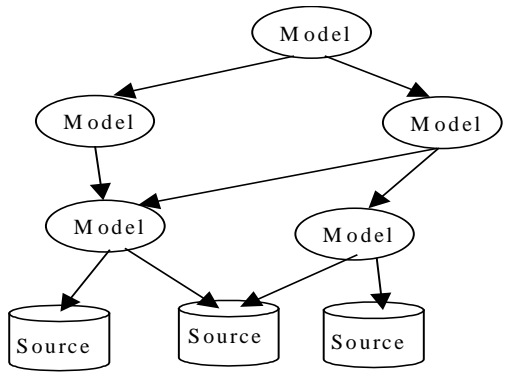

**Figure 2. Model dependencies**

If a user poses a query using a certain model a rewrite algorithm will retrieve the specifications of other models needed to retrieve the data. The model dependencies form an acyclic graph. The rewrite algorithm follows the paths starting at the initial model to find all integration models involved.

## 3. Model Specifications

In Quilt, model specifications are used to return XML documents. Within the models these tags are used to change the "semantic" meaning to the data. The following shows an example for the MuonParts (Muon parts are parts of the CMS Muon sub detector and stored in a separate database):

FUNCTION MUONPart()

{

FOR $a IN DOCUMENT("MUON.xml")/MPart

RETURN

<Part name=$a/NameDescription/@name

ID=$a/@id Weight=$a/@weight/>

}

The schema of the Muon database (MUON.xml) contains "<Mpart>" tags while the result returns "<Part>" tags and changes the attribute "NameDescription" into "name", thereby changing the meaning of the data from a local to a more global context. The MUONPart function can now be used as a separate model or used as part of an integration model that will use the "<Part>" tag.

Figure 3 shows an example in which ECALPart and MUONPart are two models that are used within a model about parts and channels (channels are used to transform analog signals from the parts they are related to into digital signals. Information about channels is stored in a separate data source.)

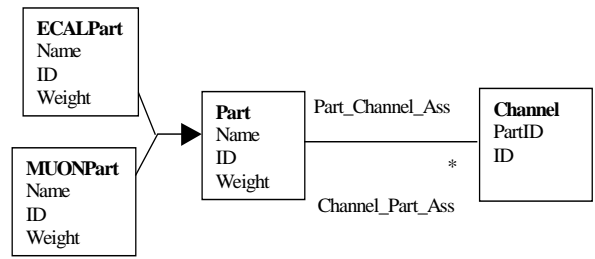

**Figure 3 Integration models**

The associations between a part and a channel are also specified in Quilt. The following is an example of the Part_Channel_Ass association where $a is a part element:

FUNCTION Part_Channel_Ass($a)

{

  FOR  $b IN Channels()

  WHERE

    $a/@ID=$b/@PartID

  RETURN

    $b

}

Users can now pose the following query: Return the channels of parts with a weight less than 100:

FOR $a IN Parts()

WHERE

   $a/@Weight<100

RETURN

  Part_Channel_Ass($a)

For the prototype reported in this paper a so-called global as view approach is followed i.e. models are defined as functions of the source. Within data integration there are generally two approaches: the global as view and local as view [16]. The reason for choosing a global as view approach is that it is straightforward to implement. Furthermore, the current models can be expressed using a global as view (although this is not always possible [16]).

## 4. Management of Models

In time the structure, location and semantics of sources can change. This can result in a change within the integration models. It is not always possible to anticipate what this change will be and updating integration models will therefore not be completely automatic. However, based on assumptions it can be partly automated. The updates are based on the structure of the global as view models described in the previous section. Two types of change are identified: syntax change and semantic change. In syntax change the data that are needed by an integration model are located in another schema or object. However the semantics of the data stays the same. Consider the following example: Suppose a model M0 returns a list (XML file) of parts containing the luminosity and material strength. These data are located in source1.xml. With Quilt this can be expressed as follows:

```
<Parts >
FOR $a IN DOCUMENT("source1.xml")//Part
RETURN
<Part Luminosity=$a/luminosity/@L
 MaterialStrength=$a/MatStr/@Strg >
</Part>
</Parts>
```

It is possible that these data are outdated and that the new data are stored in another database with a different schema. This could result in the following model M1:

```
<Parts>
FOR $a IN DOCUMENT("source2.xml")//ECALpart
RETURN
 <Part Luminosity=$a/lum/@lsty
MaterialStrength=$a/Material/@Strength>
</Part>
</Parts>
```

Other models that use this model do not have to change because the semantics of the output stays the same (Part, Luminosity, MaterialStrength). If there is versioning of models, dependent models need to create a new version. This is done automatically (Figure 4). If there is no versioning the changed model can simply be replaced without consequences for the other models.

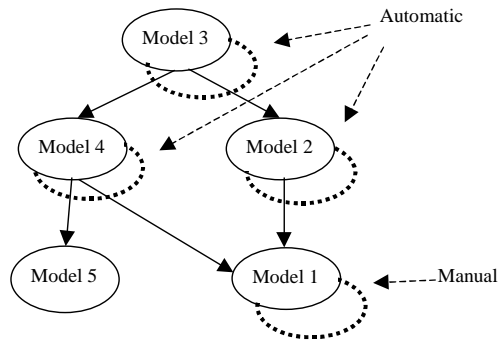

**Figure 4. Propagated versioning**

Besides syntax change, there can also be a semantic change i.e. the meaning of the model can change. This change is much more difficult to handle. How do other models use this new meaning? Does it replace models with the same meaning or is it an addition? It is often difficult to automate this kind of change.

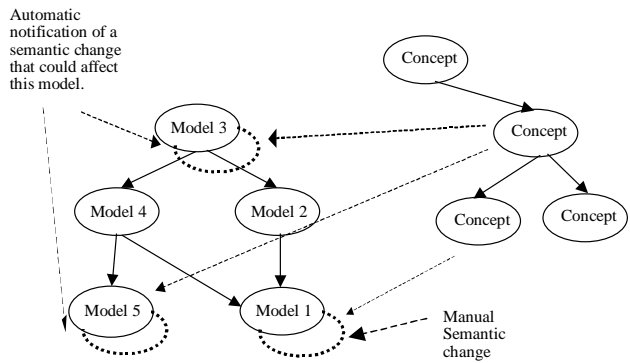

**Figure 5. Semantic change**

However, it is possible to support model engineers in managing the models when a semantic change occurs. Attached to every model there is a "meaning" of the model (i.e. the model represents certain concepts). This is a description of the model in human-understandable terms. Furthermore, this concept can be based on other concepts. When a semantic change occurs, or when a new model with certain semantics is introduced within the environment, models that are based on the semantics of the model are notified. Figure 5 shows an example of a semantic update. It is up to the model engineers to decide if the models are changed.

Currently these concepts form a taxonomy (that is a hierarchical classification of concepts). A more general approach to semantic updates would be the use of an ontology instead of a taxonomy [17]. However, the concepts used at the moment have a strong hierarchical structure.

In our prototype, the models are currently stored within an XML file. When model engineers want to develop a new model they can query this XML file to see the current model and model dependencies. Thereby getting an overview of what is available and can be reused.

# 5. Observations

Several observations can be made with respect to the current prototype. Kweelt [15] was used to describe models and queries. At that time Kweelt was the only implementation of Quilt. The implementation was straightforward and not optimized. The composition of models out of other models created complex queries. These queries could take a long time to execute depending on how such a query was specified. It is our belief that in the near future XML query engines will be more optimized.

We noticed that not every integration model could be expressed within Quilt or other query languages. One example is a matrix multiplication on a data set to perform a correction. To enable the specification of these models a query language should be extended with plug-ins that perform actions not available in standard query languages. Within Kweelt it is possible to use externally defined Java functions. However, within the current implementation this was difficult to use. Instead we defined plug-ins outside the models (as defined with Quilt). These plug-ins represent models too. However they are not expressible within the query language.

Figure 6 shows an example: A user wants to access data through the top level model that integrates data from a database and data selected within an interactive selection procedure (e.g. in a user interface selecting graphs that follow a certain curve). The data for the interactive selection came from a simulation. This simulation accessed data from a database and another integration model. Notice that the different integration models, selection and simulation could have been packed together. However, the different components can now be reused within other integration models.

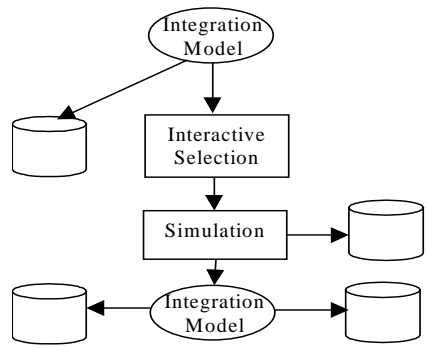

**Figure 6. Data flow**

The simulation and interactive selection models are integration models that cannot be expressed within the query language. Currently, in the prototype, the models represent "virtual" data. Data are only materialized if a user requests it. However, within certain domains, fast retrieval is important rather than having the latest updates. Instead of materializing on-the-fly, data could be materialized as needed and used multiple times (a so-called caching mechanism).

XSL [18] is a specification to create a different view on an XML document. However, there are some limitations to XSL. The output of XSL is again an XML document. Furthermore, the mechanism to select data is based on Xpath.

In [19] the next specification of DOM (DOM level 2) the notion of views is introduced. Views are XML documents with an associated stylesheet. These views can be compared with the models described in this paper and this shows several differences. The functions of our models can be specified as queries. XSL only allows Xpaths and Quilt queries have more expressive power than Xpaths. However, we believe that in due time XSL will be based on a more general selection mechanism than Xpaths. A likely candidate is a query language for XML (indeed Quilt is a proposal for this). Furthermore, we described certain management functionality not currently available for Xpaths. Notice that instead of models specified in Quilt, models could be stylesheets.

# 6. Related Work

The system described in this paper is based on mediator technology. The approach described in this paper has much in common with the TSIMMIS project [20],[21]. The approach taken in MIX [22] is similar to the approach within the TSIMMIS project. Within MIX, XML is used instead of OEM (as in TSIMMIS). These projects are based on a global as view approach only. The integration project HERMES [23] is based on a local as view approach. Within our approach we follow a global as view approach. Furthermore, we are using vendor independent standards where possible (in this case XML, Xpath and the Quilt proposal for querying) and our system is based on open source software (Kweelt [15], Xerces, Xalan [24]).

The MIX project developed its own query language, called XMAS. Furthermore, MIX focused on rapid development of mediators using XMAS. Within this prototype it is also possible to perform rapid development of mediators, but this work focuses on reuse of integration models (mediators) as part of rapid development and management of integration models.

The notion of views on semi-structured data have been discussed in [25] and the term virtual 'view' in that document relates to our models. However, nothing is said about view representation, and a hierarchy of views.

In [26] there is a description of how relations of attributes in views can be discovered using a DBMS. According to [26] this process can not be automated. However, it discusses algorithms that can help an engineer of viewpoints to make the correct relations. The algorithm is based on relational schemas but can be applied with XML as target structure (view). Their viewpoints are attributes that are functions of results of queries of attributes of some source. The algorithms exploit the structure of relational databases, and SQL. In [26] this is referred to as targeted schema mapping. Because of the structure of their viewpoints, DBMS can be used as data transformation engines.

The Agora project [27] is a project that has a similar approach to data integration with XML. Based on the Le Select [28] data integration system, it focuses on publishing relational data as XML.

[30] discusses a data integration project within bio-informatics. It uses a similar approach as [27] for data integration. However, the approach of [30] is based on the ANSI-SPARC [31] architecture. Furthermore, the wrappers reside at the mediator in [30] and the query language is based on Icode [32].

# 7. Conclusions

The goal of this first prototype was to explore the possibilities of XML for data integration and to deal with model management. Not every model can be expressed within Quilt. However, by introducing plug-ins that have an XML interface it is possible to cope with that. The approach described in this paper allows physicists (in the role of model engineers) to focus more on modeling issues. Furthermore, physicists (in the role of users) can query data using their domain model. The domain models create a transparent access for users to data in multiple sources. The model management creates transparency for the model engineers and promotes reuse of models.

The current prototype contains five simple models of which two are accessible for users. One model is used in both end user models. Furthermore, two plug-in models were used in combination with integration models constructed with Quilt. In total four different data sources were used.

Most data models used today are hardwired into applications and are not well documented. Furthermore, several data sources are not well maintained and have poor description of the schemas. This makes it difficult to describe these models in such a framework and create transparent access to data. However, it is important that there will be a framework that creates transparent access to data. Because of legal obligations certain data has to be traceable.

In the current prototype the models need to be typed in, instead of developing them with a GUI. The models described using Quilt could be mapped to models defined in Protégé [29]. Protégé is a tool to define ontologies on top of knowledge bases. The difference is that Protégé offers type safety on the models and constraint management using first order logic. Within our models we can define instances and relations as queries. Currently this is not possible within Protégé. However, Protégé offers a transparent environment for developing models that is complementary to the functionality we currently implemented. Furthermore, it is open source and extendable.

The current prototype has a centralized architecture. But within the CMS experiment there are approximately 40 data sources (lower bound) and even more users. Therefore the system has to become scalable. This could be achieved by creating a more distributed application.

Although this paper uses XML to abstract from different sources, other languages can be used too. If there are only relational databases, ODBC drivers and SQL could be used to define viewpoints. The framework relates to the ideas expressed in [10] and [11] in representing data sources as XML and creating/managing domain specific models (views) on this data.

## Acknowledgments

The authors like to thank CERN, University of the West of England and Eindhoven University of Technology for their support and Christoph Koch for his input in writing this paper.